# Tracking Human Behavioural Consistency by Analysing Periodicity of Household Water Consumption

Seán Quinn
Insight Centre for Data Analytics
Dublin City University
Dublin, Ireland
sean.quinn@insight-centre.org

Noel Murphy
School of Electronic Engineering
Dublin City University
Dublin, Ireland
noel.murphy@dcu.ie

Alan F. Smeaton
Insight Centre for Data Analytics
Dublin City University
Dublin, Ireland
alan.smeaton@dcu.ie

## ABSTRACT

People are living longer than ever due to advances in healthcare, and this has prompted many healthcare providers to look towards remote patient care as a means to meet the needs of the future. It is now a priority to enable people to reside in their own homes rather than in overburdened facilities whenever possible. The increasing maturity of IoT technologies and the falling costs of connected sensors has made the deployment of remote healthcare at scale an increasingly attractive prospect. In this work we demonstrate that we can measure the consistency and regularity of the behaviour of a household using sensor readings generated from interaction with the home environment. We show that we can track changes in this behaviour regularity longitudinally and detect changes that may be related to significant life events or trends that may be medically significant. We achieve this using periodicity analysis on water usage readings sampled from the main household water meter every 15 minutes for over 8 months. We utilise an IoT Application Enablement Platform in conjunction with low cost LoRa-enabled sensors and a Low Power Wide Area Network in order to validate a data collection methodology that could be deployed at large scale in future. We envision the statistical methods described here being applied to data streams from the homes of elderly and at-risk groups, both as a means of early illness detection and for monitoring the well-being of those with known illnesses.

## 1. INTRODUCTION

People are living longer than ever, and the complications of this changing demographic profile are starting to show in different populations around the globe. For example in Bolzano, Italy where 1 in 5 of the population are over the age of 65, there are 3 people over 65 for every 2 people under 20 and for every 2 people entering the workforce 3 people are leaving it [10]. In Bolzano and soon in many more locations around the world, the challenge has become how to provide high quality health and social services while coping with economic pressures such as fixed governmental budgets. Taking a broader global view, today 1 in 10 people are over 60 years old, by 2050 this will be 1 in 5 and over 60's will outnumber children aged 0-14 [12]. This radical societal shift is just around the corner, and collectively we will need to have the technological and health support systems in place when the predictable impact begins to occur, or the net effect will be widespread, unnecessary suffering, particularly amongst the vulnerable in society.

Many in the healthcare industry see remote healthcare as an effective remedy to the pressures of an ageing demographic with remote patient care growing by 44% in 2016 [2] and achieving promising results in terms of both cost efficiency and clinical outcomes. There have been reports of remote health care saving providers up to $150k per patient per year [7], reducing hospital admissions by up to 40% [7] and reducing readmissions by as much as 75% [4]. A study on heart failure patients reported a reduction of 50% in post-discharge chronic heart rate failure due to remote monitoring [11].

The adoption and future prospects of remote healthcare are dependent on a number of factors, including:

1. the maturity of Internet of Things (IoT) technologies,
2. the falling cost of connected sensors,
3. our ability to extract value from data, through machine learning and signal processing.

In this paper we address the third factor and explore techniques to extract potential medically significant insights from sensor data. We address this challenge by deploying sensors into a typical household to gather data on inhabitants' interactions with their home environment. From this data we aim to measure the regularity of a household occupant's behaviour, and in turn detect significant changes to this behaviour over time.

We utilise an IoT Application Enablement Platform (AEP) in conjunction with low-cost LoRa sensors [15] and a Low Power Wide Area Network (LPWAN) to harvest the data. The analysis we carry out in this paper focuses on a single metric, namely, total household water consumption, measured in litres and sampled every 15 minutes for over 8 months. We use periodicity analysis as our main tool to establish a behavioural profile and to detect changes in occupant behaviour by monitoring the strength of 12 hour and 24 hour periodicities over time.

## 2. BACKGROUND AND RELATED WORK

### 2.1 Connected Water Metering

A framework is described in [16] for regularly sampling a household water meter. In this work, the objective was disaggregation of the water usage – identifying the proportion of the overall household consumption for which each water-consuming appliance is responsible. This is comparable to deriving an itemised water bill from only the data captured at the main household meter. The water consumption breakdowns were then correlated to a number of socio-economic factors in order to establish the water usage habits of different types of households. The success of this analysis was rooted in the high sampling rate of the water meter – taking a reading every 10 seconds. This work helped to inform our own data capturing methodology.

### 2.2 Remote Monitoring of the Elderly

A framework for the detection of abnormal behaviours in elderly people with dementia through home instrumentation is presented in [13]. Data is captured from numerous door contact sensors and motion sensors placed around the home, which track the location of an individual within the home. The authors employ visualisation and clustering techniques to build a profile of normal activity patterns. Behaviour that deviates from expected patterns is primarily detected based on how long an individual is active in a given room of the home and at what time of day this activity is recorded. Our own work takes a slightly longer-term perspective in that we seek to monitor changes in the regularity of behaviour over time rather than identify particular instances of abnormal behaviour.

### 2.3 Periodicity Analysis of Sensor Data

Periodicity detection is a well-established technique used to find repeating patterns in a signal [1]. Periodicity has been successfully applied in a number of fields including engineering, astronomy, biology, and physics [17]. In this paper we focus on the application of periodicity detection to time-series sensor data and we draw on our previous work in [8], [3] and [9]. In this previous work we applied periodicity analysis to accelerometer data captured from wrist-worn smart watches. In particular, [8] demonstrated that relevant periodicities can be detected from wrist-worn sensors, and we introduced the idea of measuring the intensity of periodicities over time. We contrasted six different methods for measuring the intensity of a periodicity, giving special consideration to irregularly-sampled data, and data that is missing large numbers of values. In [3] we went one step further and showed the correlation between 24hr periodicity as detected from wrist-worn accelerometers, with cardiometabolic risk biomarkers and health-related quality of life metrics. Periodicity strength was shown to be consistently associated with LDL-cholesterol, triglycerides (a type of fat found in the blood) and High-sensitivity C-reactive Protein (hs-CRP) as well as health-related quality of life.

In [9] we showed that periodicity intensity in wrist-worn accelerometer data can be used as an effective tool to measure the success of treatments and interventions for people with sleep disorders. Our work here is similar to these three previous papers in that we also focus on calculating periodicity in sensor data streams. However, it differs in the sense that we aim to detect periodicities based upon an individual's interaction with their environment rather than by directly measuring activity through a medium such as a wearable sensor. We also seek to utilise modern IoT technologies and low-cost sensors in order to validate a data collection methodology that could realistically be deployed at large scale.

## 3. DATA GATHERING

### 3.1 Data Subjects

Given our objective of inferring household occupant behavioural patterns through observation of their water usage, consideration was given to the selection of an appropriate household and test subjects. We opted to select a household comprised solely of working adults, due to our assumption that the occupants are likely to have a structured, repetitive lifestyle and are representative of a large population segment. We are cognisant of the fact that we would likely have obtained different results had there been children or non-working adults in our test household. The household was occupied by two individuals during the first 6 months of the data gathering and by three individuals during the final 2 months.

### 3.2 Data Gathering Framework

A meter was installed on the main water supply to capture the consumption of the entire household in litres (L) and was sampled every 15 minutes. The sensor is a battery powered LoRa-enabled [15] IoT device connected to a Low Power Wide Area Network (LPWAN) through Pervasive Nation[1]. Pervasive Nation is a nationwide IoT test-bed in Ireland, and it operates a Low Power Wide Area network that provides connectivity to large parts of the country. The flow of data from the sensor was managed by Davra[2], which provides an IoT Application Enablement Platform (AEP) used to operate the Pervasive Nation network and thus enable rapid provisioning of connected sensors as well as providing real-time monitoring and access to all data gathered.

### 3.3 Duration and Data Volume

The data gathering phase of this experiment ran from the 9th of September 2017 until the 31st of May 2018 for a total duration of 8 months and 23 days (265 days). We recorded a total of 24,994 readings from the water meter during this period. Figure 1 shows the raw water meter data captured.

### 3.4 Unusual Events

We note the significance of a number of dates within the data gathering phase of this experiment that may have had

---
[1]http://connectcentre.ie/pervasive-nation
[2]http://davra.com

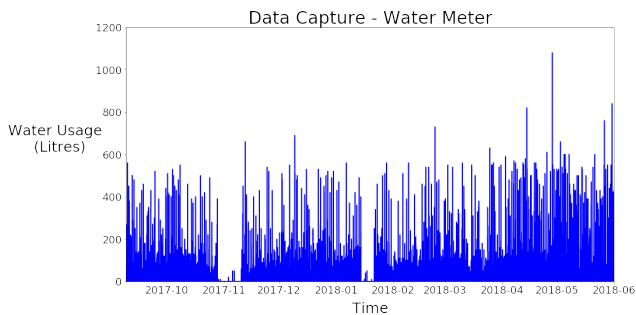

**Figure 1: Graph of data capture for the water meter. Note the two sizeable gaps where the household occupants were on vacation.**

an effect on the water usage by the household's occupants:

- Storm Ophelia: This ex-hurricane hit Ireland on 16th October 2017 and resulted in widespread cancellations of work, schools, colleges and travel. The occupants did not travel to work on this date.
- Storm Emma & Large Snowfall: The main snowfall of storm Emma lasted 4 days and took place between the 1st and 4th March 2018. It was the heaviest snowfall in the Dublin area in 36 years. The occupants did not travel to work on these dates.
- Public holidays: There were 8 public holidays during the duration of the data capturing phase. On these days the occupants did not travel to work.
- Hardware Errors: During the course of the experiment there was 5 days in which significant data loss was experienced. These outages were the result of various hardware and connectivity faults.
- Family vacations: Two family vacations took place during the duration of the data gathering; the first vacation took place in October/November 2017 and had a duration of 16 days. The second vacation took place in January 2018 and lasted 9 days.

We consider all other days within the data capture period to be normal days of activity by the household occupants.

## 4. ANALYSIS AND RESULTS

### 4.1 Data Pre-processing

Each reading recorded from our water meter consists of a time-stamp and a cumulative reading of the total water consumed in litres since the meter went live. The first step in data pre-processing is to calculate the water consumed within each 15 minute time interval by subtracting every reading from the reading directly succeeding it. We then remove all the days that we identified as producing abnormal readings. These are days when (i) the occupants were on Vacation, (ii) the occupants were not at work on a weekday due to a public holiday or weather event, or (iii) hardware issues caused significant disruptions to the readings. These abnormal days are described in section 3.4. Table 1 shows how many days of data remain for each day of the week after these days have been removed.

**Table 1: Count of Number of Days in Data**

| Day | Count |
|---|---|
| Monday | 28 |
| Tuesday | 34 |
| Wednesday | 34 |
| Thursday | 34 |
| Friday | 31 |
| Saturday | 33 |
| Sunday | 32 |

**Table 2: Readings Recorded Per Day**

| No. of Readings | Count |
|---|---|
| 96 | 36 |
| 95 | 93 |
| 94 | 58 |
| 93 | 24 |
| 92 | 15 |

Theoretically, a sensor that samples every 15 minutes results in 96 readings per day. In practice, a time delay of between 1 to 30 seconds occurred for every sample. This means we often recorded fewer than 96 reading in a day, but never more than 96 readings and never less than 15 minutes between readings. The distribution of number of readings recorded per day is depicted in Table 2.

The final step of our pre-processing is to apply data binning. We binned each day into 96 15-minute intervals, where a reading is assigned to an interval if it falls between the start and end time of that interval. Empty bins occur in days with less that 96 readings and these bins are assigned an 'n/a' value.

### 4.2 Data distribution

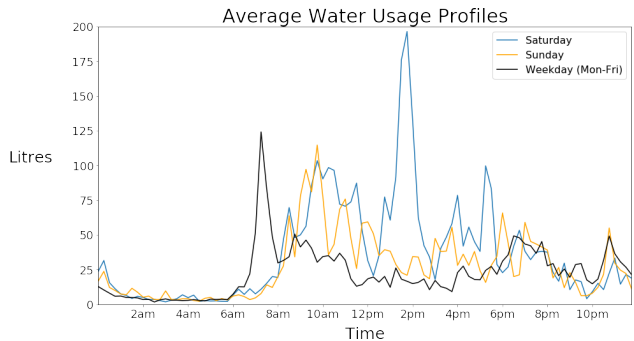

**Figure 2: Average water usage in litres at each time of day for Saturdays, Sundays and weekdays (Monday-Friday)**

We split the dataset into each respective day of the week (Monday–Sunday) and calculated a 24-hour usage profile for each day by calculating the mean for each of the 96 time bins. We observed three distinct profiles as follows: The distributions for each of the weekdays, Monday to Friday, were highly similar to each other while, Saturday and Sunday were distinct from both each other and from the weekday profile. These three profiles are shown in figure 2. The key differences between the profiles are (i) the time at which the "morning spike" occurs: 7am for weekdays vs 10am for Saturdays and Sundays, and (ii) the large 2pm spike occurring on Saturdays. It is worth noting that all profiles follow similar trends during the night-time hours of 7pm to 6am.

We further calculated the standard deviation across each of the 96 time bins for our weekday (Monday–Friday) profile.

This is shown alongside the mean in Figure 3. The standard deviation is higher than the mean at all times of the day, indicating that there is a lot of variation in our expected usage at any given time of day.

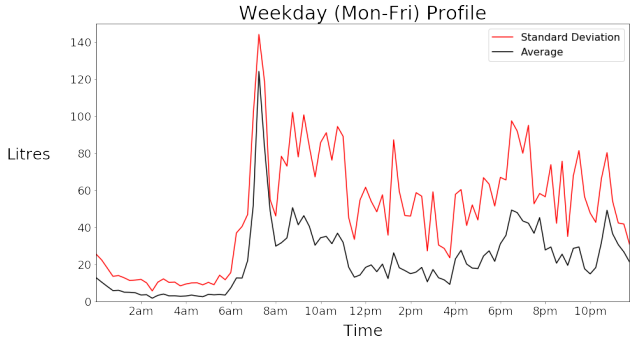

Figure 3: Average water usage and standard deviation at each time of the day for weekdays (Monday-Friday)

### 4.3 Power Spectral Density & Periodicity

Power Spectral Density (PSD) estimation is a method that can be used to detect significant repeating cycles in any kind of time-series data or signal. It does this by calculating how strong the expected signal power is at a given frequency of the input signal. A periodogram is a visualisation of the PSD for a continuous spectrum of frequencies of the input signal [1]. By calculating and visualising PSD in this way we can observe which frequencies (repeating cycles) are strong within our signal. PSD and Periodogram analysis has been used to estimate the spectrum of discrete and continuous signals in a wide variety of different scenarios and application domains [17]. In our case the signal we are observing is water usage, so we are measuring the regularity and the cycles present within the household occupants' water usage.

We know that human behaviour naturally follows a relatively stable cycle known as a circadian rhythm [6]. This is essentially a 24-hour periodicity that is driven by biological processes and further compounded by many common social constructs, such as working fixed hours every day. So by verifying that at least a 24-hour periodicity exists in our water meter data, we can be more confident that monitoring a person's interaction with their home environment through water metering is a valid approach to approximating human behavioural regularity. In our analysis, in order to establish if periodicities exist in our data and if so, whether or not they persist over time, we spilt our data into 10-day windows and calculated a periodogram for each. These results are plotted in Figure 4, where the periodograms for each of the 134 time windows have been overlaid on top of each other. We detected a strong 12-hour and 24-hour periodicity in all windows. A 10-day window size was selected as it gives sufficient data to calculate a stable periodogram while also being sufficiently sensitive to changes over time, though we also investigated other window sizes. Each 10-day window starts one day later than the previous window, so there is an overlap of nine days between directly adjacent windows.

### 4.4 Periodicity Intensity

Once we detect the existence of a periodicity in our data,

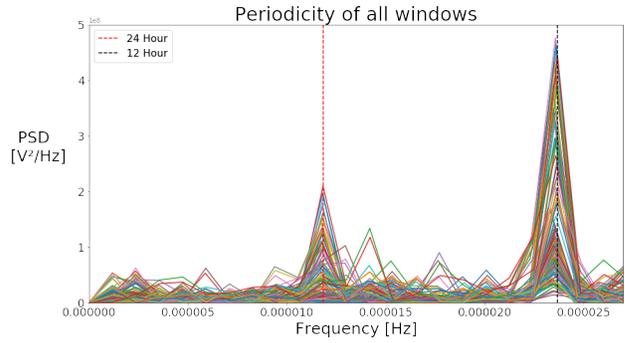

Figure 4: Periodogram for each of the 134 10-day windows (overlaid on top of each other).

we can track its intensity (strength) over time to observe whether or not the regularity of the occupants' behaviour is changing. We do this with our data by plotting the magnitude of the PSD for the frequencies corresponding to 12 and 24 hours for each window taken on our data. These results are shown in Figure 5. We also highlight on this graph the two periods where the household's occupants were on vacation. We are now in a position to observe changes in the regularity of the occupants' behaviour, and potentially explain these changes with information that we have about events occurring in the occupants' lives. An unexpected insight revealed by this analysis is that our occupants behaviour gradually becomes more irregular in the lead-in to a vacation, and also takes time to gradually become regular again upon return. This finding was consistent with oral interview with one of the occupants, where they noted "vacation preparations" and "settling back in afterwards" as disruptions to their daily routines.

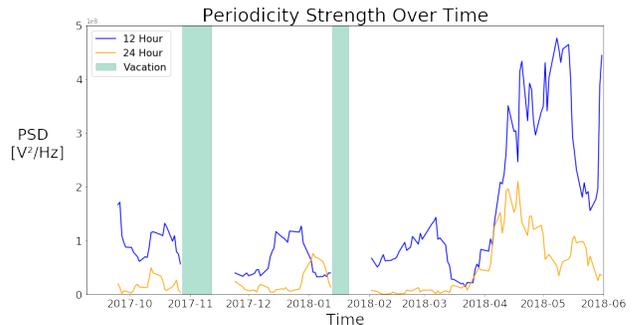

Figure 5: Strength of the 12 hour and 24 hour periodicties changing over time. We also mark the two intervals where the household's occupants were on vacation.

### 5. CONCLUSIONS

In this work we have shown that we can successfully detect relevant periodicities from a household water meter connected to an LPWAN using a low-cost sensor. We have shown that variation in the strength of detected periodicities can be tracked over time by calculating periodicities on local windows of data. We observed an interesting correlation between periodicity intensity and a life event of the household occupants through the changes detected in water usage directly before and after a vacation.

The implications of this work in the context of remote health care lie in the fact that we have achieved success in tracking human behavioural regularity over time and in a non-intrusive way. Our method does not require participants to explicitly interact with any technology or record any information. We gather potentially medically significant insights without placing the burden of active participation on subjects.

Many serious illnesses impact the regularity with which people can live their lives. We postulate that symptoms such as the sleep disruptions typical of Parkinson's disease [14] and repetitive behaviours typical of Alzheimer's disease [5] may cause long term perturbations in behavioural periodicities, which, if true, could in turn be monitored using this type of analysis.

Here we presented our results on a household whose occupants have no known illnesses affecting behavioural periodicity. Based on this preliminary success our future work envisions the application of this analysis at scale to elderly and at-risk groups in a larger study informed by and carried out in conjunction with medical researchers.

We believe that the analysis and findings presented here are not limited to data captured from a water meter, but could similarly be applied to data captured from any sensed aspect of the home environment with which a person regularly interacts. Examples of other suitable candidates include room occupancy sensors, sound intensity sensors and electricity meters fitted to lights and appliances.

## 6. ACKNOWLEDGMENTS

This work is supported by Science Foundation Ireland under grant numbers SFI/12/RC/2289 and 16/SP/3804 and is co-funded under the European Regional Development Fund.